\documentclass[aps,rmp,twocolumn,superscriptaddress,showpacs]{revtex4}

\usepackage{graphicx}
\usepackage{epsfig}

\begin{document}


\pagenumbering{arabic}

\title{{Unusual Resonators: Plasmonics, Metamaterials, and Random Media}}

\author{Konstantin Y. Bliokh}
\affiliation{Frontier Research System, The Institute of Physical and Chemical Research
(RIKEN), Wako-shi, Saitama 351-0198, Japan} \affiliation{Institute of Radio Astronomy, 4
Krasnoznamyonnaya St., Kharkov 61002, Ukraine}
\author{Yury P. Bliokh}
\affiliation{Frontier Research System, The Institute of Physical and Chemical Research (RIKEN),
Wako-shi, Saitama 351-0198, Japan} \affiliation{Department of Physics, Technion-Israel Institute of
Technology, Haifa 32000, Israel}
\author{Valentin Freilikher}
\affiliation{Department of Physics, Bar-Ilan University, Ramat-Gan 52900, Israel}
\author{Sergey Savel'ev}
\affiliation{Frontier Research System, The Institute of Physical and Chemical Research (RIKEN),
Wako-shi, Saitama 351-0198, Japan} \affiliation{ Department of Physics, Loughborough University,
Loughborough LE11 3TU, United Kingdom}
\author{Franco Nori}
\affiliation{Frontier Research System, The Institute of Physical and Chemical Research
(RIKEN), Wako-shi, Saitama 351-0198, Japan} \affiliation{Center for Theoretical Physics,
Department of Physics, Applied Physics Program, Center for the Study of Complex Systems,
The University of Michigan, Ann Arbor, Michigan 48109-1040, USA}

\begin{abstract}

Superresolution, extraordinary transmission, total absorption, and localization of
electromagnetic waves are currently attracting growing attention. These phenomena are
related to different physical systems and are usually studied within the context of
different, sometimes rather sophisticated, approaches. Remarkably, all these seemingly
unrelated phenomena owe their origin to the same underlying physical mechanism -- wave
interaction with an open resonator. Here we show that it is possible to describe all of
these effects in a unified way, mapping each system onto a simple resonator model. Such
description provides a thorough understanding of the phenomena, explains all the main
features of their complex behavior, and enables to control the system via the resonator
parameters: eigenfrequencies, Q-factors, and coupling coefficients.
\end{abstract}


\maketitle
\tableofcontents

\section{Introduction}
Left-handed materials, plasmon-polariton systems, and localized modes in random media are
attracting nowadays the ever-increasing interest of physicists and engineers. This is due
to both the fundamental character of the problems and promising applications in
photonics, subwavelength optics, random lasing, etc. There have been a number of separate
investigations and reviews of these phenomena (see, e.g., \cite{Bliokh:2004a, Pendry3,
Veselago:2006b, Eleftheriades, Zayats, Ozbay, Genet, Abajo, Maier, FG, LGP,
Sheng,Shalaev}), but no detailed comparison has been attempted in spite of their deep
underlying similarities. Discussing analogies between these systems and with their simple
classical counterparts provides a more unified understanding, new insights, and can be
illuminating. There are numerous examples of fundamental physical phenomena that can be
explained in terms of classical oscillators. Resonators provide the next-step
generalization revealing additional common features such as, e.g., the ``critical
coupling effect'' crucial for open wave systems with dissipation. Below we briefly
describe the complex systems mentioned above and analyze them in terms of simple
resonator models. Note that here we do not consider microwave and optical resonators,
quantum dots, Mie resonances, impurity zones in periodic structures, cavities in photonic
srystals, etc. Our goal is to demonstrate that a broad variety of physical phenomena in
systems that do \textit{not} contain conventional resonant cavities, nonetheless can be
adequately described in terms of classical resonators.

\subsection{Veselago--Pendry's ``perfect lens''.}

In 1968 Veselago examined electromagnetic wave propagation in a virtual medium with
simultaneous negative permittivity and permeability \cite{Veselago:1967a}. He showed that
such left-handed medium (LHM) was characterized by an unusual negative refraction: the
incident and refracted beams at the interface between the LHM and ordinary media
(hereafter, the vacuum) lie on the same side of the normal to the interface. This
property implies that a flat LHM slab can act as a lens forming a 3D image of the object,
as illustrated in Fig.~1. Interest in LHM grew very fast after Pendry's paper
\cite{Pendry:2000a}, where it was shown that a LHM slab can act as a \textit{perfect}
lens. Namely, a LHM slab with permittivity and permeability having the same absolute
value as in the surrounding medium $(\varepsilon=\mu=-1)$ forms a perfect copy of an
object: all details of the object, even smaller than the wavelength of light, are
reproduced (for reviews, see, e.g., \cite{Bliokh:2004a, Veselago:2006b, Pendry3,
Eleftheriades,Shalaev}). In practice, left-handed materials are artificial periodic
structures (metamaterials), and any ``perfect lens'' will have a finite resolution
limited by the size of the unit cell.

\begin{figure}[tbh]
\centering \scalebox{0.27}{\includegraphics{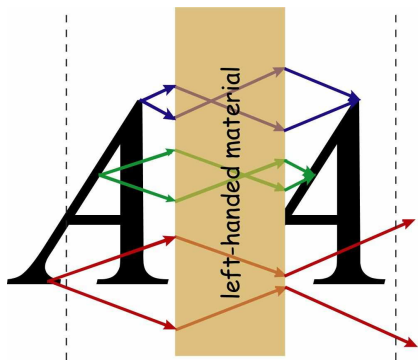}} \caption{(color
online). A flat slab of left-handed material can act as a lens forming a perfect 3D image
of any object located at a distance less than the slab thickness from the surface. In all
our figures, we denote metals in grey color, dielectrics in blue, and left-handed media
in yellow. \label{Fig1}}
\end{figure}

\subsection{Extraordinary optical transmission.}

Metallic thin films can also provide super-resolution for the near evanescent field
\cite{Pendry:2000a, Nano}. But for propagating waves, a metallic layer acts as a very
good mirror: only an exponentially small part of the radiation can penetrate through it.
Surprisingly, in 1998 Ebbesen \textit{et al.} \cite{Ebbesen1} found that an optically
opaque metal film perforated with a periodic array of \textit{sub-wavelength}-sized holes
was abnormally \textit{transparent} for certain resonant frequencies or angles of
incidence, Fig.~2. The energy flux through the film can be orders of magnitude larger
than the cumulative flux through the holes when considered as isolated (for reviews, see,
e.g., \cite{Genet, Abajo, Zayats}). In addition to its fundamental interest, this effect
offers promising applications as tunable filters, spatial and spectral multiplexors, etc.
(see, e.g., \cite{Sambles, Lezec}).

\begin{figure}[tbh]
\centering \scalebox{0.265}{\includegraphics{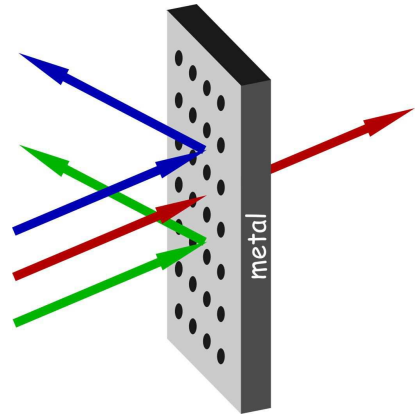}} \caption{(color
online). Resonant transparency of a perforated metal film. A periodically modified
(perforated or corrugated) optically thick metal film becomes essentially transparent for
certain resonant frequencies or angles of incidence.} \label{Fig2}
\end{figure}

\subsection{Total absorption of electromagnetic waves.}

Total internal reflection (TIR) occurs when an oblique light beam strikes an interface
between two transparent media, and the refractive index is smaller on the other side of
the interface. For instance, the incident light is totally reflected from the prism
bottom (which is the TIR surface), as shown in Fig.~3a. A polished silver plate is also a
very good mirror that reflects all the incident light, as in Fig.~3b. However, when the
TIR surface and the plate are located right next to each other, the reflected beam can
disappear and all the light can be \textit{totally absorbed} by the silver plate
\cite{Otto}, as illustrated in Fig.~3c.

Total absorption can also be observed in the microwave frequency band. When replacing the
prism by a reflecting sub-wavelength diffraction grating, Fig.~3d, and the silver plate
by an overdense plasma (i.e., a plasma whose Langmuir (plasma) frequency is higher than
the incident wave frequency), Fig.~3e, the same effect appears: the incident
electromagnetic wave can be totally absorbed by the plasma \cite{Bliokh4, Wang}, Fig.~3f,
even though both elements act separately as very good mirrors.

\begin{figure}[h]
\centering \scalebox{0.42}{\includegraphics{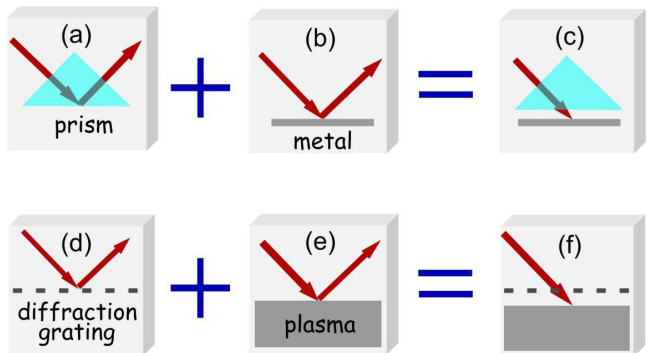}} \caption{(color
online). The total absorption of electromagnetic waves in optics (a,b,c) and plasma
physics (d,e,f). Strikingly, even though elements (a,b) or (d,e) act separately as very
good mirrors, their combination can absorb all the incident radiation (c,f).}
\label{Fig3}
\end{figure}

\subsection{Localized states.}

Extraordinary optical transmission and total absorbtion can be observed in a quite
different system: 1D random dielectric media. Although the medium is locally transparent,
the wave field intensity typically decays exponentially deep into the medium, so that a
long enough sample reflects the incident wave as a good mirror. This is because of
multiple wave scattering in randomly-inhomogeneous media producing a strong (Anderson)
localization of the wave field \cite{Anderson} (for reviews, see, e.g.,
\cite{LGP,FG,Sheng}). A simple manifestation of this effect is the almost-total
reflection of light from a thick stack of transparencies \cite{Berry}. However, there is
a set of resonant frequencies, individual for each random sample, which correspond to a
\emph{high transmission} of the wave through the sample accompanied by a large
concentration of energy in a finite region inside the sample \cite{Azbel,Azbel2} (see
Fig.~4). Like optical ``speckle patterns'', such resonances (localized states) represent
a unique ``fingerprint'' of each random sample. In active random media, regions that
localize waves are sources of electromagnetic radiation producing a so-called random
lasing effect, which offers the smallest lasers, just a few-wavelengths in size
\cite{Wiersma,Cao,Milner}. If the sample has small losses, the resonant transparency can
turn into a \emph{total absorption} of the incident wave \cite{Bliokh3}.

\begin{figure}[tbh]
\centering \scalebox{0.41}{\includegraphics{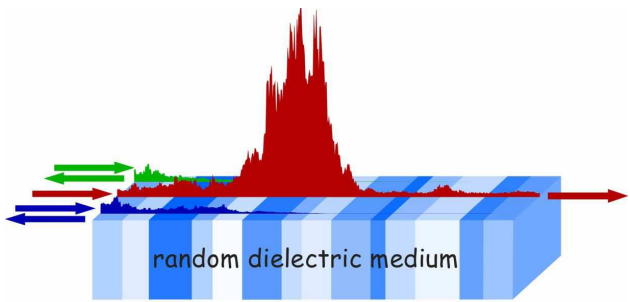}}\caption{(color online).
Resonant wave transmission through a 1D random dielectric sample. The spatial
distributions of the intensity of the resonant (red) and non-resonant (green and blue)
waves are depicted on the top of the sample which is displayed schematically.}
\label{Fig4}
\end{figure}

\section{Classical resonators}

\subsection{Basic features.}

The notion of resonator implies the existence of eigenmodes {\it localized} in space. The
localization of modes is usually achieved by a sandwich-type ``mirror-cavity-mirror''
structure which is analogous to a quantum-mechanical potential well bounded by potential
barriers and can be of any nature (see Fig.~5). In a closed resonator without dissipation
each mode is characterized by its resonant frequency (energy level) $\omega_{\rm res}$
and spatial structure of the field $\chi(\textbf{r})$. The eigenmode field $\Psi$ can be
factorized as
\[ \Psi(\textbf{r},t)=\psi(t)\,\chi(\textbf{r})~,\]
where $\psi$ is a solution of the harmonic oscillator equation:
\[\frac{d^2\psi}{dt^2}+\omega_{\rm res}^2\psi=0~.\]
Depending on whether the modes are localized in
all spatial dimensions or not, the resulting spectrum can be either discrete or
continuous.

The resonator can be non-conservative due to internal \emph{dissipation} of energy.
Furthermore, the barriers can allow small energy \emph{leakage} either from or to the
cavity, e.g. due to `under-barrier' tunnelling via evanescent waves. In such cases one
has to consider the resonator as an open system with quasi-modes characterized by fuzzy
energy levels of a finite width, Fig.~5. The time dependence of the fields is not purely
harmonic anymore and can be described as an oscillator with damping:

\begin{equation}
\label{eq1} \frac{d^2\psi}{dt^2}+\omega_{\rm res}Q^{-1}\frac{d\psi}{dt}+\omega_{\rm
res}^2\psi=0~.
\end{equation}
Here, the dimensionless \textit{Q-factor} characterizes the total losses in the
resonator:

\begin{equation}
\label{eq2} Q^{-1}=Q^{-1}_{\rm diss}+Q^{-1}_{\rm leak}\ll 1~,
\end{equation}
where $Q_{\rm diss}$ and $Q_{\rm leak}$ are the Q-factors responsible for the dissipation
and leakage, respectively. The dimensionless half-width of the resonant peak in the
spectrum equals \[\delta\nu \equiv \frac{\delta\omega_{\rm res}}{\omega_{\rm res}} =
Q^{-1}~.\]

\begin{figure}[tbh]
\centering \scalebox{0.47}{\includegraphics{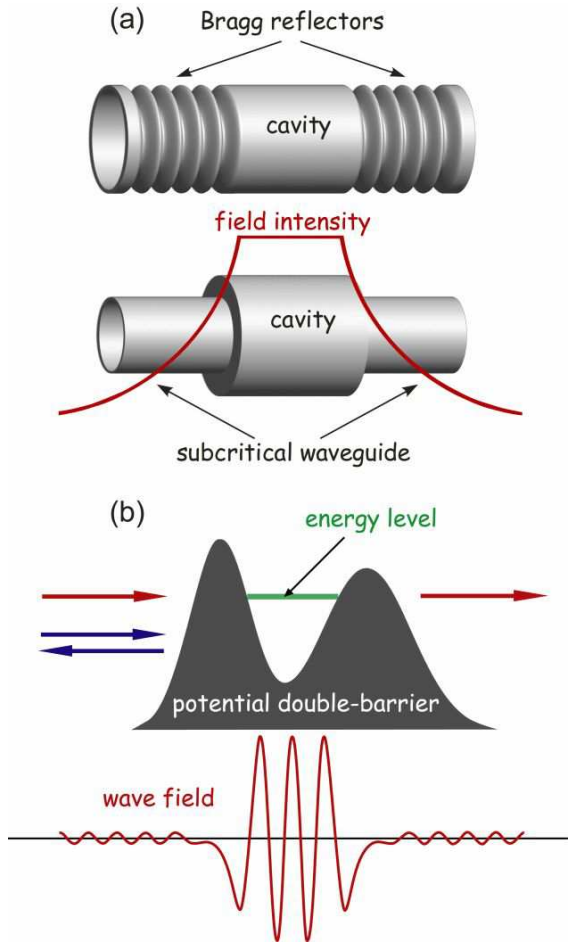}} \caption{(color
online). Examples of classical and quantum open quasi-1D resonators. (a) A waveguide
segment (cavity) is surrounded by Bragg-reflecting or subcritical segments (acting as
barriers). The field inside the resonator can interact with an external wave field
through non-propagating evanescent modes in the barriers. A 3D generalization of the top
system can be a cavity inside a photonic crystal in the frequency gap. (b) Any potential
well can represent a quantum resonator. Open resonators are surrounded by finite-width
barriers. An incident particle can effectively tunnel through both energy barriers when
its energy coincides with one of the energy-levels in the cavity \cite{Bohm}. A
characteristic quasi-mode wave function is depicted at the bottom. } \label{Fig5}
\end{figure}

\subsection{Plane wave interacting with a resonator.}

The tunnelling of an incident plane wave through an open 1D resonator is characterized by
the transmission and reflection coefficients $T$ and $R$. The transmittance $T$ is
usually small due to the opaque barriers, but if the frequency of the incident wave
coincides with one of the eigenmode frequencies, an effective resonant tunneling occurs.
The corresponding transmission coefficient, $T_{\rm res}$, is given by \cite{Bohm, Yariv,
Bliokh3}:

\begin{equation}\label{eq3}
T_{\rm res}={4Q_{\rm leak\, 1}^{-1}Q_{\rm leak\, 2}^{-1}\over\ \left( Q_{\rm leak\,
1}^{-1}+Q_{\rm leak\, 2}^{-1}+Q^{-1}_{\rm diss}\right)^2}~.
\end{equation}
Here $Q_{\rm leak\, 1}$ and $Q_{\rm leak\, 2}$ are the leakage Q-factors ($Q_{\rm
leak}^{-1}=Q_{\rm leak\, 1}^{-1}+Q_{\rm leak\, 2}^{-1}$), which are related to the
transmittances $T_1$ and $T_2$ of the two barriers by \cite{Bliokh4}

\begin{equation}\label{eq4}
Q_{\rm leak\, 1,2}^{-1}=\frac{v_g T_{1,2}}{2\ell\omega_{\rm res}}~.
\end{equation}
Here $v_g$ is the wave group velocity inside the resonator and $\ell$ is the resonator
cavity length (so that $2\ell /v_g$ is the round-trip travel time of the wave inside the
cavity). Hereafter we will assume $\omega_{\rm res}/v_g=k$, where $k$ is the wave number
of the resonant wave. Note that the total transparency, $T_{\rm res}=1$, is achieved only
in a dissipationless symmetric resonator, i.e.

\begin{equation}\label{eq5}
T_{\rm res}=1~~{\rm when}~~Q_{\rm diss}^{-1}=0~,~~Q_{\rm leak\, 1}^{-1}=Q_{\rm leak\,
2}^{-1}~.
\end{equation}

The reflection coefficient $R$ is close to unity off-resonance and is characterized by
sharp resonant dips on-resonance. The resonant reflection coefficient is given by
\cite{Bohm, Yariv, Bliokh3}:

\begin{equation}\label{eq6}
R_{\rm res}={\left( -Q_{\rm leak\, 1}^{-1}+Q_{\rm leak\, 2}^{-1}+Q^{-1}_{\rm
diss}\right)^2\over\ \left( Q_{\rm leak\, 1}^{-1}+Q_{\rm leak\, 2}^{-1}+Q^{-1}_{\rm
diss}\right)^2}~.
\end{equation}
In contrast to the transmittance (\ref{eq3}), the reflectance (\ref{eq6}) reaches its
minimum value also in dissipative asymmetric resonators \cite{Yariv, Bliokh3}:

\begin{equation}\label{eq7}
R_{\rm res}=0~~{\rm when}~~Q_{\rm diss}^{-1}=Q_{\rm leak\, 1}^{-1}-Q_{\rm leak\,
2}^{-1}~.
\end{equation}
This is the so-called \textit{critical coupling} effect.
In the important particular case when the second barrier is opaque, $Q_{\rm leak\,
2}^{-1}=0$, $Q_{\rm leak\, 1}^{-1}\equiv Q_{\rm leak}^{-1}$, so that the total
transmittance vanishes, $T\equiv 0$, the reflectance spectrum exhibits pronounced
resonant dips, with $R_{\rm res}=0$, if the leakage and dissipation Q-factors are equal
to each other (see, e.g. \cite{Slater}): $Q_{\rm diss}^{-1}=Q_{\rm leak}^{-1}$. Then, the
incident wave is totally absorbed by an open resonator, so that all the wave energy
penetrates into the resonator and dissipates therein.

\subsection{Coupled resonators.}

Two resonators can be coupled by the fields penetrating through the barriers. This system
(outside the critical coupling regime) can be effectively described by the coupled
oscillators model. When the first (incoming) resonator is excited by a monochromatic
source with frequency $\omega$, the appropriate oscillator equations can be written as
follows:
\begin{eqnarray}\label{eq8}
{d^2 \psi_{\rm in}\over d\tau^2}+{Q^{-1}}{d\psi_{\rm in}\over d\tau}+\psi_{\rm
in}=q\psi_{\rm
out}+f_0e^{-i\nu \tau}~,\nonumber\\
{d^2 \psi_{\rm out}\over d\tau^2}+{Q^{-1}}{d\psi_{\rm out}\over d\tau}+\psi_{\rm
out}=q\psi_{\rm in}~,
\end{eqnarray}
where $\psi_{\rm in}$ $(\psi_{\rm out})$ is the field in the first (second) resonator,
$\tau=\omega_{\rm res}t$ and $\nu=\omega/\omega_{\rm res}$ are the dimensionless time and
frequency, $q\ll 1$ is the coupling coefficient, and $f_0$ is the effective exciting
force from the incident field. The steady-state solutions of Eqs.~(\ref{eq8}) are
oscillations with amplitudes $A_{\rm in}$ and $A_{\rm out}$ given by:
\begin{eqnarray}\label{eq9}
A_{\rm in}={f_0\left(1-\nu^2-i\nu Q^{-1}\right)\over
\left(1-\nu^2-i\nu Q^{-1}\right)^2-q^2}~,\nonumber\\
A_{\rm out}={f_0q\over\left(1-\nu^2-i\nu Q^{-1}\right)^2-q^2}~.
\end{eqnarray}

Near-resonance, $|\nu-1|\ll 1$, the frequency dependencies of these amplitudes at
different values of the $qQ$ factor are shown in Fig.~6. There can be seen that when the
condition $qQ>1$ is satisfied, there are two collective resonant modes with equal field
amplitudes in the two resonators. Their frequencies are shifted from the eigenfrequency
of the oscillators, $\nu =1$, due to losses and coupling:
\begin{equation}\label{eq10}
\nu_{\rm res}^{\pm}=1\pm \frac{1}{2}\sqrt{q^2-Q^{-2}}~.
\end{equation}
As $qQ$ decreases, the resonant peaks in the spectra are located near each other and meet
when $qQ=1$. In the regime $qQ<1$ there is one peak at $\nu=1$.

The parameter $qQ$ that appears in the model has a simple physical meaning: it determines
whether the two resonators should be considered as essentially coupled or isolated. When
$Q^{-1}\ll q$, the losses are negligible and the field characteristics are essentially
determined by the coupling.  Remarkably, in this case the field intensity in the first
(incoming) resonator is negligible at $\nu=1$, and almost all the energy is concentrated
in the second resonator: $A_{\rm out}\gg A_{\rm in}$. On the contrary, when the losses
prevail over the coupling, $Q^{-1}\gg q$, the incident wave only excites the first
resonator, and the energy is concentrated mostly in it: $A_{\rm in}\gg A_{\rm out}$.
\footnote[1]{All these properties can be easily seen in our animated simulations at
http://dml.riken.jp/resonators/resonators.swf where three regimes mimicking perfect
lenses, enhanced transparency and weak coupling (see the next Section) are illustrated.}

\begin{figure}[h]
\centering \scalebox{0.40}{\includegraphics{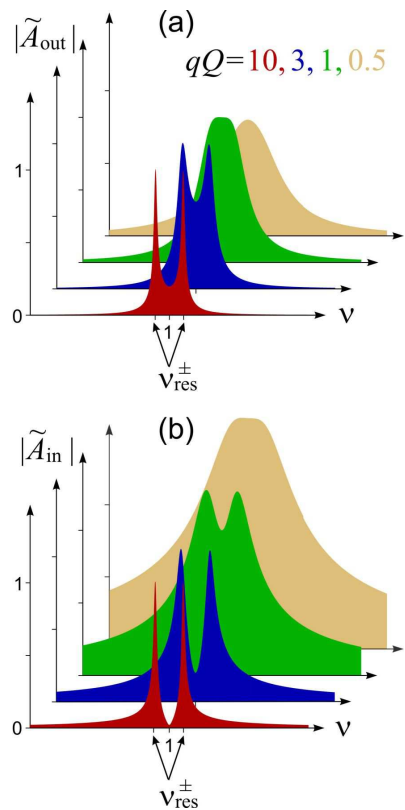}} \caption{(color
online). Near-resonant transmission of an incident wave through two coupled open
resonators at different values of $qQ$. The normalized (i.e., multiplied by the factor
$2Q^{-1}$) absolute values of the field amplitudes in two resonators, $|\tilde A_{\rm
out}|$ (a) and $|\tilde A_{\rm in}|$ (b), are shown. In the dissipationless case ($Q_{\rm
diss}^{-1}=0$, $Q=Q_{\rm leak}$) the transmission coefficient of the system is given by
$T=|\tilde A_{\rm out}|^2$.} \label{Fig9}
\end{figure}

\section{Surface plasmon-polariton systems}

\subsection{Basic features.}

The interface between materials with different signs of the permittivity, $\varepsilon>0$
and $\varepsilon<0$, (or permeability, $\mu>0$ and $\mu<0$), supports surface $p$- (or
$s$-) polarized waves, also known as \textit{plasmon-polaritons} (PP). PP were discovered
in 1957 by Ritche \cite{Ritche} while studying a metal-vacuum interface. Renewed interest
in plasmon-polaritons comes from their considerable role in contemporary nano-physics
\cite{Barnes, Zayats}. The interface between regular ($\varepsilon>0$, $\mu>0$) and
left-handed ($\varepsilon<0$, $\mu<0$) materials supports PP with an arbitrary
polarization \cite{Ruppin}.

PP are electromagnetic waves which are trapped at the interface, their electromagnetic
fields decaying exponentially deep into both media (Fig.~7a). Hence, the interface forms
a peculiar resonator with eigenmodes which are localized along the normal to the
interface but can propagate freely along the interface. Spatial eigenfunctions of this
resonator have the form:

\[\chi(\textbf{r})=\exp(-\kappa_z |z|)\exp\left(
i\textbf{k}_{\bot}\textbf{r}_{\bot}\right)~,\] and are characterized by a dispersion
relation $\textbf{k}_{\bot}=\textbf{k}_{\bot}^{\rm PP}(\omega)$. Here the interface is
associated with the $z=0$ plane, the subscript ``$\bot$'' indicates vectors within the
$(x,y)$ plane, and the decay constant $\kappa_z$ can take different values in the two
media. The plasmon-polariton resonator possesses all the features that are inherent to
usual resonators: eigenfrequency, Q-factor, topography of eigenmode fields, etc. It will
be shown below that the identification of PP as a resonator is more than an analogy,
since it captures the main physical process underlying this phenomenon.

An exponential profile of the PP field suggests that it can interact with
\textit{evanescent}, non-propagating fields from an external source, which are
characterized by a purely imaginary wave-vector component: $k_z=\pm i\kappa_z$.
Furthermore, an incident \textit{propagating} electromagnetic plane wave cannot excite
the PP resonator. This is because for propagating waves (PW) $k_{\bot}^{\rm
PW}<\omega/c$, whereas for plasmon-polaritons $k_{\bot}^{\rm PP}>\omega/c$. At the same
time, evanescent waves (EW) are characterized by $k_{\bot}^{EW}>\omega/c$ and can excite
the PP resonator. There are methods to convert a propagating wave into an evanescent one.
This allows the interaction of light with plasmon-polaritons, which has lead to a new
branch of physics: \textit{plasmonics} (see, e.g., \cite{Zayats, Ozbay, Maier}).

\subsection{Enhanced transparency of a metal film.}

Let us examine the transmission of a plane wave through an optically thick metal film
perforated with small, sub-wavelength holes, as shown in Fig.~2. Two surfaces of the
smooth metal film can be associated with two identical PP resonators, coupled by their
fields, as shown in Fig.~7b. As it has been noticed, PPs cannot be directly excited by
the incident plane wave; however, PPs can interact through periodic modulations of the
surface \cite{Tan, Bonod, Darmanyan, Dykhne}. The effective interaction of PPs with light
occurs at some resonance frequency $\omega=\omega_{\rm res}$ (or angle of propagation
$\alpha=\alpha_{\rm res}$), at which their wave vectors differ by a reciprocal lattice
vector $\textbf{K}$:

\begin{equation}\label{eq11}
\textbf{k}_{\bot}^{\rm PP}(\omega_{\rm res})-\textbf{k}_{\bot}^{\rm PW}(\omega_{\rm
res})=\textbf{K}~.
\end{equation}
One can say that the PP wave vector shifted by a reciprocal lattice vector acquires a
real $z$-component, or vice versa, the shifted propagating mode becomes evanescent in the
$z$ direction. Thus, the periodic structure can be considered as a \textit{mode
transformer}. (It is worth remarking that while in microwave electronics such structures
are used for slowing down the eigenmodes, here we deal with a speedup of the PP
eigenmodes.) The periodic corrugation of the metal surface can be treated as a
diffraction grating placed on the surface. The grating can be located at a distance $d$
from the smooth surface; it remains coupled with plasmons-polaritons by their field
\cite{Bliokh2, Lin}, as shown in Fig.~7b,c.

Now we can make use of the general theory of resonators. Assuming the dissipation to be
negligible, $Q_{\rm diss}=0$, the Q-factor of the PP resonator is determined by the
energy leakage from the resonator due to the transformation of the evanescent waves into
the propagating ones:

\begin{equation}\label{eq12}
Q_{\rm leak}^{-1}=\gamma\exp\left(-2\kappa_z d\right)~.
\end{equation}
Here, $\gamma\ll 1$ is the transformation coefficient at the diffraction grating, and the
exponential dependence arises due to decay of the evanescent field intensity between the
metal surface and grating. The coupling coefficient between the two PP resonators is
determined by the field of the first resonator acting on the second one:
\begin{equation}\label{eq13}
q=\exp\left(-\kappa_{z}\Delta\right)~,
\end{equation}
where $\Delta$ is the film thickness. The dependence of the transmission coefficient $T$
on the normalized incident wave frequency $\nu$ is described by
Eqs.~(\ref{eq8})--(\ref{eq10}), and is illustrated in Fig.~6a. It exhibits two nearby
peaks: $T_{\rm res}=1$ at $\nu=\nu_{\rm res}^{\pm}$, when $Q_{\rm leak}^{-1}<q$, or one
peak $T_{\rm res}<1$ at $\nu=1$ when $Q_{\rm leak}^{-1}>q$. A characteristic field
distribution for the total transmission is shown in Fig.~7b. The coupling parameter
decreases as $\Delta$ grows, and the transmission spectrum $T(\nu)$ changes as in
Fig.~6a. The same dependence of the transmission spectrum on the film thickness has been
obtained in \cite{Tan, Dykhne, Martin-Moreno1, Benabbas} and in many other papers
concerned with particular geometries and specific modifications. In fact, all these
features are just general properties of two coupled resonators, independently of details.

Thus, the light transmission through a perforated (corrugated) metal film can be divided
into three processes: (i) transformation of the incident plane wave into an evanescent
wave on the first diffraction grating, (ii) resonant ``penetration'' of the evanescent
field through two coupled plasmon-polariton resonators, (iii) reverse transformation into
the propagating wave on the second grating, Fig.~7b. It may seem at first that the larger
the transformation coefficient at the diffraction grating is, the better the transmission
is. However, larger transformation coefficients result in smaller Q-factors. When
$\gamma$ exceeds a critical value, the transmission becomes less than one and decreases
with $\gamma$ \cite{Dykhne}.

\subsection{Critical coupling in optics and plasma physics.}

In the above model, it is easy to incorporate dissipation characterized by a small
imaginary part of the dielectric constant,
$\varepsilon=\varepsilon^\prime+i\varepsilon^{\prime\prime}$, by introducing the
dissipation Q-factor

\begin{equation}\label{eq14}
Q_{\rm diss}^{-1}\ =\
\frac{\varepsilon^{\prime\prime}}{|\varepsilon^{\prime}|}\ \ll\
1~.
\end{equation}
The dissipation is negligible only if $Q_{\rm diss}^{-1}\ll \min\left(Q_{\rm
leak}^{-1},q\right)$. Otherwise, even very small dissipation will drastically affect the
resonance phenomena as it is compared to the exponentially small parameters (\ref{eq12})
and (\ref{eq13}). If $Q_{\rm diss}^{-1}\sim Q_{\rm leak}^{-1}$, the dissipation may
substantially suppress the transmission. When $Q_{\rm diss}^{-1}\geq q$, the dissipation
breaks down the coupling between two PP resonators on either side of the film, and they
can be regarded as essentially independent. For the system under consideration this means
that only the first resonator will be excited by the incident wave and the metallic film
can be considered as a semi-infinite medium, Fig.~7c.

\begin{widetext}

\begin{figure}[tbh]
\centering\scalebox{0.85}{\includegraphics{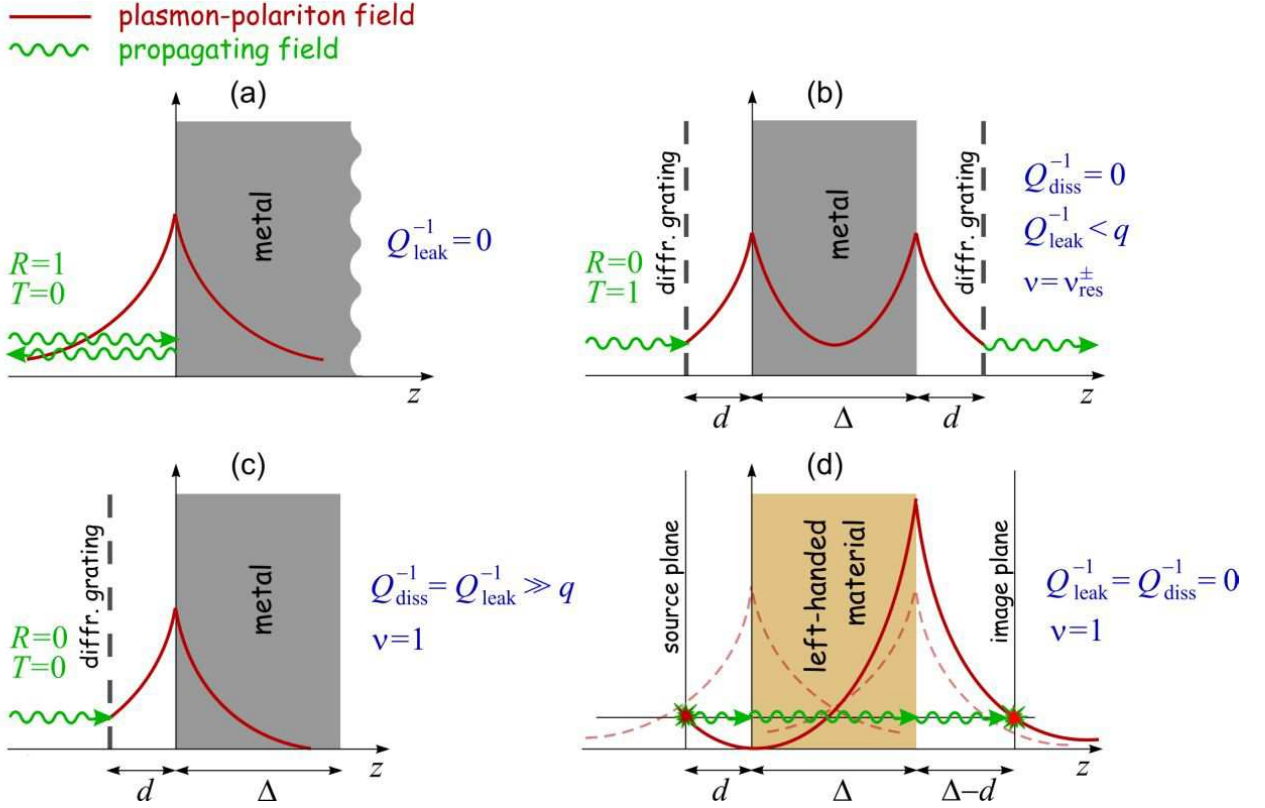}}\hspace{20mm}
\caption{(color online). Schematic diagrams of surface plasmon-polariton (PP) resonators,
their interaction with incident waves, and the corresponding resonators parameters. (a) A
vacuum/metal interface supports plasmon modes localized in the $z$-direction and reflects
propagating waves. Propagating and evanescent waves do not interact with each other, and,
therefore, there is no energy leakage from the PP resonator. (b) A metal film represents
a system of two PP resonators coupled by their fields. A diffraction grating (or total
internal reflection (TIR) interface, Fig.~3) can transform a propagating wave into an
evanescent wave and vice versa, thereby making the PP resonators open, $Q_{\rm
leak}^{-1}\neq 0$. A resonant total transparency can be achieved when the dissipation is
negligible, whereas the coupling is strong enough, cf. Figs.~2 and 6. (c) In the critical
coupling regime all the incident wave is absorbed by the metal or plasma due to intrinsic
dissipation, cf. Fig.~3. (d) A slab of an ideal LHM also represents two coupled PP
resonators. There are two essential distinctions as compared to metal films: (i) A LHM is
transparent for propagating fields and (ii) plasmon-polaritons are always in resonance
with evanescent fields from the source. The latter means that the PP field distribution
corresponds to the $\nu =1$ point in Fig.~6, and all the evanescent field energy is
concentrated at the output surface in the dissipationless case. As a result, both
propagating and evanescent fields form an exact copy of the source field in the focal
point, cf. Fig.~1.} \label{Fig7}
\end{figure}

\end{widetext}

In such a case, the transmission through the film vanishes at all frequencies. At the
same time, the resonances show up in the reflection spectrum which exhibits sharp dips at
some frequencies. For some critical distance between the diffraction grating and metal,
$d=d_{\rm c}$, the resonant reflectance drops to zero, $R_{\rm res}=0$, and the incident
wave is totally absorbed by the metal, Fig.~3. This effect can be readily explained in
terms of the same resonator model: (i) the incident plane wave transforms into an
evanescent mode at the diffraction grating and (ii) it excites a PP resonator at the
metal surface and is totally absorbed due to the critical coupling effect, Fig.~7c. In
our case the critical coupling condition, Eq.~(\ref{eq7}), reads $Q_{\rm diss}^{-1}=
Q_{\rm leak}^{-1}$ with Eqs.~(\ref{eq12}) and (\ref{eq14}). The application of
diffraction gratings for the PP resonator excitation is most convenient in plasma
experiments. When a properly designed grating is placed in front of the plasma surface,
the reflected wave vanishes (see Fig.~3 bottom) \cite{Bliokh4, Wang}.

An analogous phenomenon in optics is known as ``frustrated total internal reflection''
\cite{Otto}. Similarly to a diffraction grating, a total internal reflection interface
can be used for plasmon-polariton excitation on a metal surface \cite{Otto, Kretschmann}.
In the so-called Otto-configuration, a metal is placed at a distance $d$ from the bottom
of a prism where the light is totally reflected (top of Fig.~3). The incident light, with
$k_{\bot}^{\rm PW}<\omega/c$, penetrates into the prism with the wave vector projection
$n\;k_{\bot}^{\rm PW}>\omega/c$ ($n>1$ is the refractive index of the prism). It
generates an evanescent wave in vacuum near the bottom and can excite the the PP
resonator at a resonance frequency $\omega=\omega_{\rm res}$ (or angle of propagation
$\alpha=\alpha_{\rm res}$) where

\begin{equation}\label{eq15}
\textbf{k}_{\bot}^{\rm PP}=n\,\textbf{k}_{\bot}^{\rm PW}~.
\end{equation}
The leakage Q-factor of the PP resonator is given by Eq.~(\ref{eq12}), where $\gamma\sim
1$ is the coefficient of transformation to the evanescent wave at the bottom of the
prism. At a critical distance $d=d_{\rm c}$ the reflected light disappears, which
evidences the critical coupling regime. Note that when the metal film is so thin that $q
\geq Q^{-1}$, the high resonant transparency of the film can be observed when the
identical prism is located symmetrically near the opposite side of the film
\cite{Dragilla}. This configuration is absolutely analogous to the above-considered
grating--metal--grating system, Fig.~7b.

The total absorption of an incident wave due to the critical coupling can be used for the
simultaneous determination of both the real and imaginary parts of the metal (plasma)
permittivity. On the one hand, the resonant frequency $\omega_{\rm res}$ (or the angle of
incidence) is determined by the resonance with the PP mode, which depends on the real
part of the metal permittivity, $\varepsilon^\prime$. On the other hand, the critical
coupling distance $d_{\rm c}$ between the prism (grating) and the metal (plasma) surface
depends on the dissipative Q-factor (\ref{eq14}) related to the imaginary part
$\varepsilon^{\prime\prime}$ of the permittivity. Thus, the critical coupling regime
provides a mapping between $(\omega_{\rm res},d_{\rm c})$ and
$(\varepsilon^\prime,\varepsilon^{\prime\prime})$ and makes it possible to retrieve the
complex permittivity of the metal (plasma) via external measurements.

\subsection{Superresolution of LHM lenses.}

While a dielectric medium is transparent for propagating plane waves and a metal surface
supports PP evanescent modes, a left-handed material combines both of these features. Let
the source of a monochromatic electromagnetic field (the object) be located at a distance
$d$ from the surface of a flat slab of an `ideal' LHM ($\varepsilon=\mu=-1$) of width
$\Delta>d$, as shown in Fig.~7d. The source radiates propagating plane-wave harmonics
with $k_{\bot}^{\rm PW}\leq\omega/c$, as well as evanescent waves with

\begin{equation}\label{eq16}
k_{\bot}^{\rm EW}>\frac{\omega}{c}~.
\end{equation}

The propagating waves are focused by the LHM slab and form the image on the opposite side
of the slab, Fig.~1. The ideal LHM is perfectly matched with the vacuum due to the
equivalence of their impedances $Z=\sqrt{\mu/\varepsilon}$, and, therefore, there is no
reflected wave in this case. The image field is almost equal to the source one: all the
plane waves reach the focal plane (located at a distance $\Delta-d$ from the slab) with
the same phase as they had in the source plane. The abberation (imperfection) of the
image might only be caused by the loss of evanescent harmonics, which are responsible for
the sub-wavelength details of the object.

Remarkably, even sub-wavelength information is not lost in the ideal LHM (Pendry's) lens.
This can be easily understood if we consider the surfaces of the LHM slab as two coupled
PP resonators, as we did to explain the high transmission of the perforated metalic
films. Evanescent waves from the source excite the first PP resonator. A distinguishing
feature of the PP mode at the ideal LHM/vacuum interface is that its dispersion relation
is precisely the same as for evanescent modes in the vacuum \cite{Ruppin}. This implies
that \textit{all} the evanescent waves (\ref{eq16}) are in resonance with the PP
resonator on the ideal LHM surface \cite{Haldane}. In other words,
$\omega\equiv\omega_{\rm res}$ and $\nu\equiv1$ for any frequency (the material
dispersion is neglected here).

The evanescent field distribution can be found from Eqs.~(\ref{eq8}) and (\ref{eq9}). The
PP resonators at the dissipationless LHM surface are characterized by an infinite
Q-factor,

\begin{equation}\label{eq17}
Q_{\rm leak}^{-1}=0~,
\end{equation}
because there is no leakage from the PP to radiative modes. We associate the amplitudes
$A_{\rm in}$ and $A_{\rm out}$ with the field amplitude at the input and output surfaces
of the slab. Then the effective external force is given by

\[f_0=A_0 \exp(-\kappa_z d)~,\]
where $A_0$ is the amplitude of the evanescent field of the source. The coupling
parameter is given by Eq.~(\ref{eq13}), as it was for the metallic film. According to
Eqs.~(\ref{eq9}) with $\nu=1$ and $Q^{-1}=0$ (see also Fig.~6), the input resonator is
not excited,

\[|A_{\rm in}|=0~,\]
whereas the amplitude at the output is exponentially large:

\[|A_{\rm out}|=\frac{|f_0|}{q}=|A_0| \exp[-\kappa_z (d-\Delta)]~.\]
In the image half-space, the evanescent field decreases with the same decrement
$\kappa_z$ and at the distance $\Delta-d$ from the second interface (the total distance
from the source is $2\Delta$) takes on the initial value (Fig.~7d):

\[|A(2\Delta)|=|A_0|~.\]

Since the phases of evanescent waves are not changed along the $z$-axis, the evanescent
fields in the focal plane precisely reproduce those in the source plane. This means that
the image created by both propagating and evanescent waves is a \textit{perfect} copy of
the source. Exactly the same evanescent field distribution in the Pendry's lens follows
from the accurate solution of Maxwell equations (see, e.g., \cite{Haldane, GomezSantos}).
It is worth noting that the electromagnetic nature of waves has not been involved in our
consideration of sub-wavelength imaging. The same result can be achieved using other
kinds of waves, e.g., liquid-surface waves \cite{Hu}, surface electromagnetic waves
propagating along special types of interfaces~\cite{kats,Shadrivov}, and surface
Josephson plasma waves~\cite{yamp} in layered superconductors.

If a small dissipation is present in LHM, it can be taken into account by introducing the
dissipation Q-factor (ref{eq14}) in Eqs.~(\ref{eq8}) and (\ref{eq9}) (here, for
simplicity, the permeability is assumed to be real). The destructive effect of
dissipation in the LHM, reducing the image quality, is defined by the ratio between the
Q-factor and the coupling parameter $q$. When $Q_{\rm diss}^{-1}\ll q$, the image
abberation is small. When $Q_{\rm diss}^{-1}\geq q=\exp(-\kappa_z \Delta)$, the
dissipation is crucial and practically destroys the penetration of evanescent waves
through the LHM slab. This limitation of the ideality of Pendry's lens has been discussed
in \cite{Garcia, Nieto-Vesperinas} using a wave approach. At the same time, the
dissipation affects the propagating waves in a LHM lens in the same manner as in normal
media, because the LHM slab do not form resonator for propagating waves. The dissipation
significantly affects the propagation waves only when $Q_{\rm diss}^{-1}\geq
(k\Delta)^{-1}\gg q$.

\begin{widetext}

\begin{figure}[tbh]
\centering \scalebox{0.58}{\includegraphics{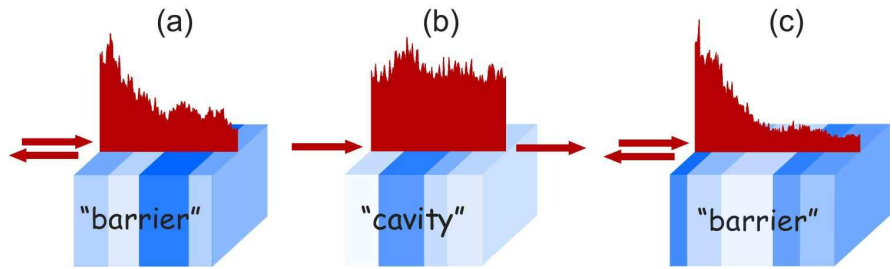}}
\caption{(color online). The sample in Fig.~4 is now ``cut'' into three separate segments
which are considered independently. It can be seen that while the right- and left-side
segments are practically opaque due to Anderson localization, the central part (where a
huge energy concentration has been observed in resonance) happens to be almost
transparent for the resonant frequency. This provides the standard
``barrier--cavity--barrier'' resonator structure, which explains the resonant features of
the sample at a given frequency.} \label{Fig10}
\end{figure}

\end{widetext}

\section{Random media}

\subsection{Resonant tunnelling.}

The resonant transmission of waves through a 1D random sample is accompanied by a large
concentration of energy inside the sample, as shown in Fig.~4. Such field distributions
can be regarded as quasi-modes of an open system. Among various localized states, high
transparency accompanies only those modes that are located near the center of the sample.
Localized modes and anomalous transparency can be explained by the existence of effective
high-Q resonator cavities inside of the sample. Figure~8 demonstrates the transparencies
of different parts of the sample from Fig.~4. It is clearly seen that the middle section,
where the energy was concentrated, is almost transparent to the resonant frequencies,
while the side parts are practically opaque to the wave. Thus, each localized state at a
frequency $\omega=\omega_{\rm res}$ can be associated with a typical resonator structure
comprised of an almost transparent segment (``cavity'') bounded by essentially
non-transparent regions (``barriers'') \cite{Bliokh1}. The wave tunnelling through such a
system can be treated as a particular case of the general problem of the transmission
through an open resonator. The distinguishing feature of the random medium is that there
are no regular walls (the medium is locally transparent in each point) and transmittances
of barriers are exponentially small as a result of Anderson localization. Moreover,
different segments of the sample turn out to be transparent for different frequencies,
i.e., each localized mode is associated with its own resonator.

The resonant tunnelling through a random sample can be described by Eq.~(\ref{eq3}),
where the barrier transmittances and Q-factors are estimated as \cite{Bliokh1}
\begin{equation}\label{eq18}
T_{1,2}\simeq \exp\left(-\frac{2L_{1,2}}{\ell_{\rm loc}}\right)~,~~Q_{\rm leak\,
1,2}^{-1}=\frac{T_{1,2}}{2k\ell}~,
\end{equation}
and the total leakage Q-factor is $Q_{\rm leak}^{-1}=Q_{\rm leak\, 1}^{-1}+Q_{\rm leak\,
2}^{-1}$. Here $L_{1}$ and $L_{2}$ are the distances from the cavity to the ends of the
sample (the barrier lengths) and $\ell_{\rm loc}$ is the localization length. The latter
is the only spatial scale responsible for the localization, and the cavity size should be
estimated as $\ell\sim\ell_{\rm loc}$. This simple model drastically reduces the level of
complexity of the problem: the disordered medium with a huge number of random elements
can be effectively described now through a few characteristic scales, namely, the
localization length, the wavelength, and the size of the sample. All the main features
and characteristics of the resonances (transmittance, field intensity, spectral
half-width and the density of states) can be estimated from the resonator model in good
agreement with experimental data \cite{Bliokh1}. In particular the perfect resonant
transmission takes place only for a symmetric dissipationless resonator, Eq.~(\ref{eq5}),
which corresponds to wave localization in the middle of the sample and a maximal total
Q-factor.

Note that a long enough sample can contain two or more isolated transparent regions where
the wave field is localized. These form the so-called ``necklace states'' predicted in
\cite{Lifshits2, Pendry3} and observed in \cite{Wiersma2, Genack}. Necklace states can be
easily incorporated in our general scheme as two or more resonators coupled by their
evanescent fields, Eqs.~(\ref{eq8})--(\ref{eq10}) \cite{unpublished}. The coupling
coefficient between the nearest resonators is

\begin{equation}\label{eq19}
q\simeq\exp\left(-\;\frac{\Delta}{\ell_{\rm loc}}\right)~,
\end{equation}
where $\Delta$ is the distance between the effective cavities.

\subsection{Critical coupling.}

It seems reasonable to assume that dissipation in the sample material worsens the
observability of resonances. However, surprisingly, small losses can improve the
conditions for the detection of localized states. A large number of resonances, which are
not visible in the transmission spectrum, become easily detected in the
\textit{reflection} spectrum \cite{Bliokh3}, Fig.~9. This is clarified in terms of the
critical coupling phenomenon. If the dissipation, $Q_{\rm diss}^{-1}$, given by
Eq.~(\ref{eq14}), exceeds the leakage, $Q_{\rm leak}^{-1}$, for the modes localized in
the middle of the sample:

\[Q_{\rm diss}^{-1}>\frac{1}{k\ell_{\rm loc}}\exp\left(-\frac{L}{\ell_{\rm loc}}\right)~,\]
the transmission is strongly suppressed for all frequencies, and $T\ll 1$. At the same
time, the other states located closer to the input of the sample and, therefore,
characterized by higher $Q_{\rm leak}^{-1}$ can be excited. According to the critical
coupling condition (\ref{eq7}), the resonant reflectance drops to zero when the
dissipation and leakage Q-factors are of the same order. For the modes localized in the
first half of the sample we can set, with an exponential accuracy, $Q_{\rm leak\,
2}^{-1}\simeq 0$, $Q_{\rm leak\, 1}^{-1}=Q_{\rm leak}^{-1}$, so that the critical
coupling condition reads $Q_{\rm diss}^{-1}=Q_{\rm leak}^{-1}$ as for the case of a
semi-infinite medium.

\begin{figure}[tbh]
\centering \scalebox{0.36}{\includegraphics{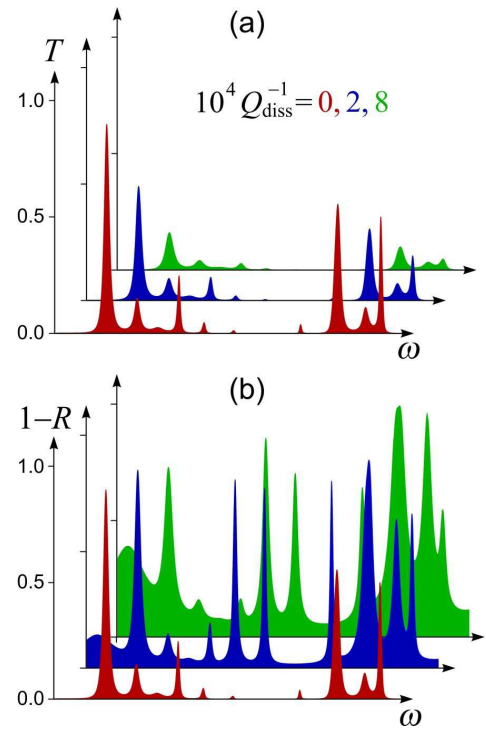}}
\caption{(color online). Spectra of the transmittance $T$ (a) and the reflectance $R$ (b)
for various values of the dissipation rate $Q_{\rm diss}$ in a random dielectric sample.
Although the dissipation is extremely small, peaks of resonant transmittance disappear
rapidly when the dissipation increases. At the same time, peaks in $1-R$ become sharply
pronounced and become even more informative than in the dissipationless case. Resonances
with $R=T=0$ evidence the critical-coupling regime.} \label{Fig12}
\end{figure}

The resonator model enables one to find characteristic parameters of localized states and
of the sample via external measurements of the transmission and reflection coefficients
\cite{Bliokh1,Bliokh3,scales}. By measuring resonant and typical off-resonance values of
the transmittance and reflectance, Eqs.~(\ref{eq3}) and (\ref{eq6}), along with the
resonance spectral half-width, it is possible to determine (at least, to estimate) the
localization length, dissipation factor, position of the localized state, and its field
intensity. Some of these internal quantities usually cannot be determined via direct
measurements, but are crucial, e.g., for the random lasing problem. For example, the
critical coupling condition connects the position of the localization region with the
dissipation rate in the medium, while the latter can be determined through the half-width
of the resonant deep in the reflectance. It should be also noted that random systems
consisting of repeated elements of several types can exhibit transmission resonances of
another kind, which are not accompanied by the energy localization and cannot be
described by the resonator model, see, e.g.,
\cite{Hendricks,Nori3,Nori1,Nori2,Nori4,Nori5}.

\begin{widetext}

\hspace{5mm}
\begin{table}
\begin{tabular}{||p{4.0cm}||p{2.3cm}|p{2.3cm}|p{2.3cm}|p{2.3cm}|p{3.5cm}||}
\hline \begin{center}\textbf{Resonator characteristics}\end{center} & \begin{center}\textbf{Enhanced transparency}\end{center} & \begin{center}\textbf{Total absorption in plasma}\end{center} & \begin{center}\textbf{Frustrated TIR in optics}\end{center}& \begin{center}\textbf{Ideal LHM lens}\end{center} & \begin{center}\textbf{Localization in random medium}\end{center}\\
\hline  \begin{center}\textbf{Dissipation Q-factor,} $Q_{\rm diss}^{-1}$\end{center} & \multicolumn{5}{|p{11.4cm}||}{ \begin{center}$\varepsilon^{\prime\prime}/|\varepsilon^\prime|$\end{center}} \\

\hline \begin{center}\textbf{Leakage Q-factor,} $Q_{\rm leak}^{-1}$\end{center} & \multicolumn{3}{|p{6.9cm}|} {\begin{center}$\gamma\exp(-2\kappa_z d)$\end{center}} & \begin{center}$ 0 $\end{center} &\begin{center}$\left(T_1+T_2\right)/2k\ell_{\rm loc}$, $T_{1,2}=\exp(-2L_{1,2}/\ell_{\rm loc})$\end{center} \\

\hline\begin{center}\textbf{Coupling coefficient,} $q$\end{center} & \begin{center}$\exp(-\kappa_z\Delta)$\end{center} & \multicolumn{2}{|p{4.6cm}|} {\begin{center}---\end{center}} & \begin{center}$\exp(-\kappa_z\Delta)$\end{center} & \begin{center}$\exp(-\Delta/\ell_{\rm loc})$\end{center} \\

\hline\begin{center}\textbf{Resonance condition, $\omega=\omega_{\rm res}$}\end{center} &
\multicolumn{2}{|p{4.6cm}|} {\begin{center}$\textbf{k}_{\bot}^{\rm
PP}-\textbf{k}_{\bot}^{\rm PW}=\textbf{K}$\end{center}} &
\begin{center}$\textbf{k}_{\bot}^{\rm PP}=n\;\textbf{k}_{\bot}^{\rm PW}$\end{center} &
\begin{center}$k_{\bot}^{\rm EW}>\omega/c$\end{center} & \begin{center}
$\omega=\omega_{\rm res}$\end{center}\\

\hline
\end{tabular}
\caption{Mapping between the classical resonator characteristics and the parameters of
various physical systems discussed in this work.}
\end{table}

\end{widetext}

\section{Concluding remarks}

The sub-wavelength resolution of a flat LHM lens, abnormal transparency of a
perforated metal film, localized states in disordered media, frustrated total
internal refraction, and total absorption of an electromagnetic wave by an
overdense plasma are all phenomena related to different areas of physics and
are characterized by different spatial scales, from the nano-scale to
centimeters and larger. In spite of such enormous differences, the main
properties of these phenomena have much in common with each other and, on a
deeper level, with simple resonator systems. As we have shown, all these
phenomena can be treated in a universal way as wave transmission through one or
two coupled resonators. A careful mapping between the resonator and the problem
parameters allows one to understand thoroughly the physical properties of the
problem and forecast how the parameters affect the result.

Of course, accurate descriptions of wave transmission through complex systems (for
example, periodically-perforated metal films or random media) involve particular details
of a given sample and depend, e.g., on the geometry of the periodic structure or on the
specific realization of the random process. Nonetheless, fundamental features of these
systems which are independent of details can be revealed only through a unified approach
emphasizing the physical essence of the problem. Resonator models provide such an
approach. In some cases (e.g., for evanescent fields in the LHM lens) resonator
description results in the \textit{exact} solution of the problem. Furthermore, in the
case of random media such model is the \textit{only} formalism which enables one to
estimate the parameters of the individual localized states.

One of the important common features of all resonator systems is their high sensitivity
to internal dissipation. Even very small dissipation, which has negligible effects for
usual propagating waves, may dramatically modify the localized resonance states. Usually
dissipation destroys the resonant transmission but develops resonant dips in the
reflection spectrum. The reason for this is that the excitation of a resonator is always
accompanied by a huge field intensity therein, proportional to the Q-factor. The energy
dissipated per unit time is determined by the product of the dissipation rate and the
field intensity. Thus, a small dissipation rate is multiplied by a high Q-factor and can
be crucial. Also, establishing a one-to-one correspondence with classical resonator
allows retrieving the \textit{internal} characteristics of an investigated system using
the \textit{external} response to a probing signal (incident wave). In this way, one can
determine the complex permittivity of metal or plasma, location and field intensity of
localized states in random medium, etc. Remarkably, dissipation can be favorable for such
purposes, revealing some hidden internal features through the critical coupling effect.

To conclude, the central result of this brief review, i.e. the mapping between the
resonator and the problems' parameters, is summarized in Table~1. We have considered only
a few, probably more intriguing and non-trivial, systems allowing the resonator
consideration. A more complete list of such systems would be much longer, since any wave
system with localized modes can be treated as a generalized resonator. In particular,
numerous quantum systems (not considered here) with potential wells, tunnelling, and
relaxation could be effectively treated within the open-resonator framework (see, e.g.,
\cite{today,rakhman,Lazarides,Rakhmanov}). Finally, it should be also noted that for some
of the systems considered above there exist alternative \textit{ad hoc} methods of
description. For instance, negative refraction and optical cloaking allow a natural
representation in the geometrical formalism of general relativity \cite{Leonhardt}.

\section*{Acknowledgments}
We gratefully acknowledge partial support from the National Security Agency (NSA),
Laboratory Physical Science (LPS), Army Research Office (ARO), National Science
Foundation (NSF) grant No. EIA-0130383, JSPS-RFBR 06-02-91200, and Core-to-Core (CTC)
program supported by Japan Society for Promotion of Science (JSPS). K.B. acknowledges
support from STCU grant P-307 and CRDF grant UAM2-1672-KK-06. S.S. acknowledges support
from the Ministry of Science, Culture and Sport of Japan via the Grant-in Aid for Young
Scientists No 18740224, the EPSRC via No. EP/D072581/1, EP/F005482/1, and ESF
network-programme ``Arrays of Quantum Dots and Josephson Junctions''.


\bibliographystyle{apsrmp}



\end{document}